# Machine learning meets mass spectrometry: a focused perspective

Daniil A. Boiko, Valentine P. Ananikov*

Zelinsky Institute of Organic Chemistry, Russian Academy of Sciences
Moscow, Russia; val@ioc.ac.ru; http://AnanikovLab.ru

## Abstract
Mass spectrometry is a widely used method to study molecules and processes in medicine, life sciences, chemistry, catalysis, and industrial product quality control, among many other applications. One of the main features of some mass spectrometry techniques is the extensive level of characterization (especially when coupled with chromatography and ion mobility methods, or a part of tandem mass spectrometry experiment) and a large amount of generated data per measurement. Terabyte scales can be easily reached with mass spectrometry studies. Consequently, mass spectrometry has faced the challenge of a high level of data disappearance. Researchers often neglect and then altogether lose access to the rich information mass spectrometry experiments could provide. With the development of machine learning methods, the opportunity arises to unlock the potential of these data, enabling previously inaccessible discoveries. The present perspective highlights reevaluation of mass spectrometry data analysis in the new generation of methods and describes significant challenges in the field, particularly related to problems involving the use of electrospray ionization. We argue that further applications of machine learning raise new requirements for instrumentation (increasing throughput and information density, decreasing pricing, and making more automation-friendly software), and once met, the field may experience significant transformation.

## Introduction

Many areas of modern science and technology require the analysis of complex molecular and hierarchical biomolecular systems.[1–3] For instance, in metabolomics, researchers analyze the presence of a large number of small molecules in biological samples.[4,5] In proteomics, many proteins can be present in the mixtures and then analyzed in depth.[6,7] Catalysis research includes a description of a large set of different catalytic species.[8] In personalized medicinal science and pharmaceutical development, the key accelerating tool is to explore molecules and their complexation in living cells.[9] Industrial production, including core chemicals as well as fine chemicals and drugs, requires efficient quality and purity control instruments.[10,11]

Mass spectrometry (MS)[12,13] is a universal key method for a range of the abovementioned cutting edge areas, including many other applications.[14–16] Some MS methods show superior compound detection quality (both sensitivity and specificity for a wide range of analytes in complex mixtures) and enable quantification and structure analysis via tandem experiments or "hard" ionization methods.[17] Generation of vast amounts (in comparison with methods such as 1D NMR, FTIR, or UV-VIS)[18] of data for complex mixtures containing hundreds and thousands of compounds is an intrinsic natural ability of a wide range of MS experiments.

For modern instruments, the amounts of data in some fields have reached a point where manual analysis becomes impractical and inefficient, depending on the specific context and analysis type. One of the solutions is to use machine learning (ML) methods — algorithms that learn relationships directly from the data without the need to describe exact steps in the decision process.[19] ML has already been widely used in fields such as drug discovery,[20] electron microscopy,[21,22] and mass spectrometry.[23–26] One of the important subfields of ML is deep learning, which is focused on studying multilayer (i.e., deep) neural networks. Their high flexibility in terms of input and output modalities significantly increases interest in the field.

The use of machine learning in mass spectrometry has come a long way. It began with applying rapidly developing pattern recognition methods to mass spectrometry data[27–29] in an attempt to harness evolving computational capabilities. However, the progress in this field was constrained by three main factors: data availability, still very limited computational resources, and algorithm development. The first two issues were addressed towards the end of the 1990s with the development of electrospray ionization and the result of continuous improvement of microprocessors, respectively. However, the application of algorithms remains a challenge. With the availability of increasingly powerful computational resources, highly capable instruments, and advanced algorithms, the field is now witnessing significant changes.

This perspective gives a focused overview of the current state of research on the intersection of machine learning and mass spectrometry. Although we attempted to cover a wide spectrum of applications, this review is by no means exhaustive and primarily focused on works involving electrospray ionization and methods related to the organic chemistry background of the authors. On the ML side, we lean towards deep learning applications. We also attempt to provide a general protocol for solving MS problems with ML. Further possible innovations in the field are the focus (particularly, development of new algorithms and mass spectra representations), and critical analysis of already existing work is provided for MS development to better suit the discovery goals.

## Mass spectrometry in the machine learning era

To trace the cutting-edge level (**Figure 1**)**,** four steps of machine learning development for applications in mass spectrometry should be considered: i) early days of MS; ii) the initial pre-ML stage; iii) the first wave of applications; iv) ML in MS stagnation; and v) the second wave of ML applications in MS, which continues right now.

Very first attempts were already interesting and encouraging. The researchers were already capable of solving noticeable problems. For instance, significant focus was given to determining the structures[27] of various compounds (including sequence analysis of oligonucleotides[28]) based on mass spectra — a problem that has not been fully solved even until now. Then, together with the development of mass spectrometry methods (particularly ionization techniques such as ESI — electrospray ionization — and MALDI — matrix-assisted laser desorption/ionization), the field started to experience very fast growth until 2010, when the overall number of papers reached a plateau or arguably even a slight decrease (marked a light orange area in **Figure 1**). One of the possible explanations is that traditional ML algorithms (such as linear models, support vector machines, decision trees and tree ensembles) were fully explored, i.e., most of them have already been applied for MS, and the field remained stagnant until the rise of deep learning. As a result, we see that the amount of conducted research has increased dramatically and continues to grow with the possibility of changing the face of the field (**Fig. 1**).

The reason why we consider the current period to be very important is that the developments in the trending directions of the two fields, which had limited intersection, have now coincided.

On the one hand, the progress of equipment development resulted in the rapid growth in the volume of mass spectrometry data. For example, large datasets on the terabyte scale are collected during MALDI imaging studies even just for peak lists with up to 2500 peaks.[30] Therefore, the period has come when new advanced and more efficient methods for analyzing spectral data are needed; otherwise, the level of data disappearance will rapidly increase, which will contribute to the impracticality of manual analysis by a person.

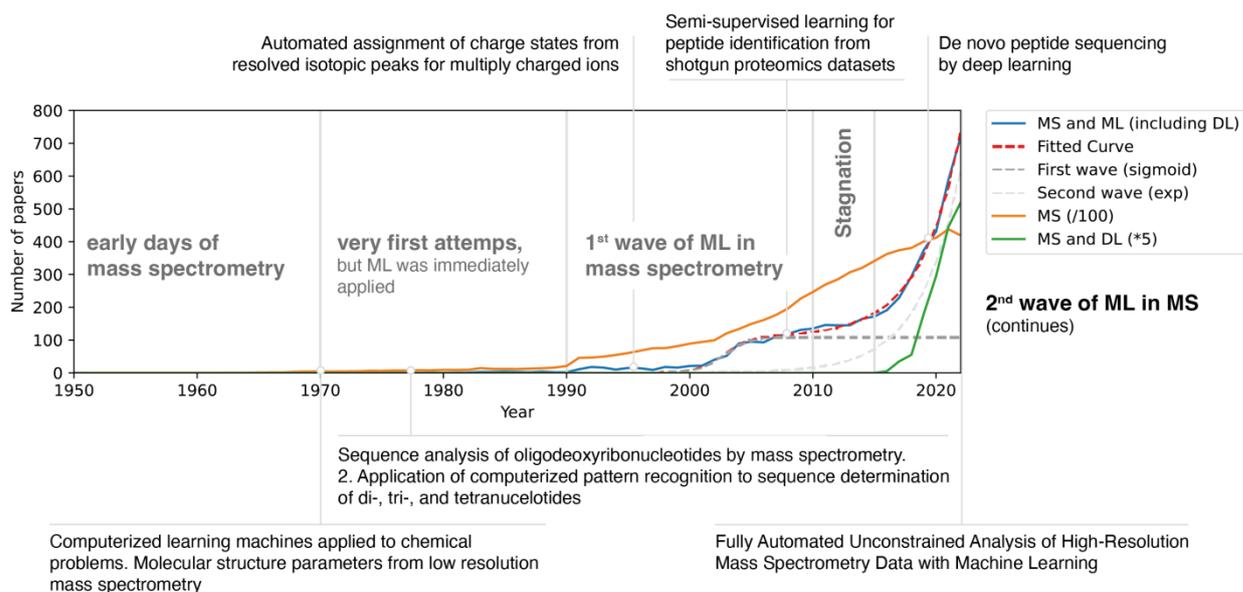

**Figure 1. Historic stages in ML-empowered MS.** The plot shows the results of the analysis of the number of papers according to Web of Science by year. We identify five periods: early days, very first attempts, first wave, ML in MS stagnation, and current period. For details, see the "Number of papers analysis" section in the Supporting Information.

On the other hand, machine learning algorithms, particularly deep learning, have reached a level that can make a qualitative change in big data analysis. Just being able to analyze the data has become a prerequisite for many fields of modern chemical and biological science, such as metabolomics. Among the reasons for this are the large size of data in terms of required computational resources and the dimensionality of data (for example, FT-ICR mass spectra may contain information about tens of thousands of ions).[31] This is additionally supported by the development of computing capabilities, including graphics processing units and distributed computations.

We have already observed significant advances caused by changes in both mass spectrometry and ML method development. Mass spectrometry is a powerful analytical technique, and when combined with machine learning methods, their synergy can provide more significant achievements, particularly in research areas that rely on mass spectrometry, such as the analysis of complex biological systems. One example is bioactive peptide discovery.[32] Mass spectrometry was also used to construct proteomics and phosphoproteomics datasets that were used to train models to predict the efficacy of anticancer drugs.[33] In another example, a machine learning pipeline trained on immune activations measured by MS was used for nanomedicine design.[34] MS data were also used without any feature extraction directly to predict lung adenocarcinoma at early stages.[35] A similar strategy enabled the detection of SARS-CoV-2.[36] Importantly, ML was employed in de novo peptide sequencing.[37]

However, the full potential of this field's overlap has yet to be seen. To date, not only have potential ML applications in MS not been fully explored (i.e., it is still hard to determine which parts of the MS research process will benefit the most from ML and how), but even the input data representation problem has not been completely addressed (i.e., it is not yet clear which representations should be used for specific tasks). Moreover, as dataset sizes increase (now a single MALDI imaging run can result in tens and hundreds of gigabytes of data), researchers face the problem of feasibility in performing calculations, which requires the development of distributed approaches. Finally, further advances in instrumentation and increased throughput will pose even greater challenges to computational work.

Below, we detail the relevant sections in this perspective to describe key trends. It is fundamental to link these two areas together and draw the attention of researchers involved in machine learning to the adaptation of algorithms for mass spectrometry problems. It is equally important to bring key equipment changes to the attention of instrument manufacturers to take full advantage of the current state of research.

# Mass spectrometry methods landscape and their potential for ML applications

In a general manner, a mass spectrometer can be described as a set of functional blocks: an inlet, an ionization source, a mass analyzer, and an ion detector (**Fig. 2a**). The system is controlled by electronics and software, and the collected data are then stored and analyzed. The types of ionization source and the analyzer have the most impact on the applicability of such data for machine learning. This is primarily due to the already advanced level of technological development in other components of a mass spectrometer.

The ionization source is the part of mass spectrometers that defines which compounds will be able to ionize and consequently enter the mass spectrometer (**Fig. 2b**). With electron ionization, the studied compound will be fragmented in the ionization chamber, and along with the molecular ion, fragment ions will be observed. The ions provide a good clue into the chemical structure of the studied compound and sometimes allow confident molecular confirmation, but for a complex system, the destructive nature of the analysis may lead to a loss of informativity. In contrast, electrospray ionization (ESI) is a "soft" ionization method that mainly provides molecular ions and, therefore, can be used to analyze the compound mixture as a whole. Observation of molecular ions allows for determining molecular formulas and significantly simplifies spectra. The simplification of spectra makes it possible to study multicomponent systems without initial separation. ESI is suitable for polar compounds (that have a dipole moment), which are common in biological samples. Despite loss of structural information, it is possible to obtain this information by employing multiple analyzers together (see below, description of the ESI-QQQ and ESI-Q-TOF instruments). Several other ionization techniques (such as desorption ionization, laser desorption/ionization, or matrix-assisted laser desorption/ionization) were also developed to make MS a universal characterization method for a variety of samples.[17]

Our ability to separate ions before detection depends on the mass analyzer (**Fig. 2c**). One of the ways to compare mass analyzers is to compare resolving powers (see

Supporting Information, "Resolution in mass spectrometry" section), defined as the ratio of *m/z* to the mass difference needed to separate two signals. Moreover, analyzers impose some limitations over mass ranges. For instance, a time-of-flight (TOF) mass analyzer shows high resolution and is often used to study large molecules, as it has a remarkably wide mass range. Instrumental progress in MS has made a range of high-performance analyzers available to achieve the desired resolution for different implementations.[17]

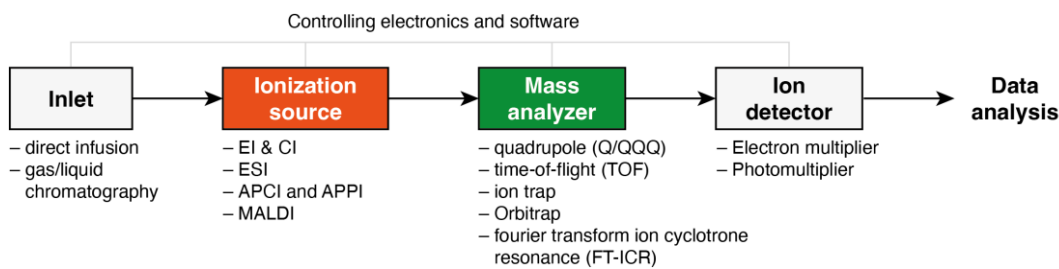

**Figure 2. Applicability of ML depending on the instrument type.** a — Diagram of a typical mass spectrometer; b — applicability of ML by the ionization method; c — applicability of ML by analyzer type; d — comparison of standard ionization method/analyzer combinations (see Supporting Information for details). The region is either the entire country or a significant part of it. By accessibility gap, we mean the difference between the number of researchers who could benefit from using these instruments and the number of those who actually have access to them. We consider applicability depending on availability.

Although these MS functional blocks can be mixed and matched in many possible combinations, some are widely used in practice (**Fig. 2d**). *EI-Q-MS* is one of the most commonly used methods to study volatile organic compounds (together with gas chromatography).[38–40] The number of these instruments is very large (approximately ten or more instruments per department; here and further, see Supporting Information for the estimations). However, the amount of obtained data is limited due to the low resolution of single-quadrupole analyzers. The main challenge here is to improve the scope of the studied compounds. Further developments in derivatization techniques will probably make the method much more appealing to machine learning algorithms by expanding the scope of ionizable compounds. *ESI-QQQ-MS* (with liquid chromatography) is a widely used method for quantitative studies but requires tedious method development (ideally, one should know not only the target but also its most characteristic fragment). Methods that combine multiple analyzers are called tandem mass spectrometry methods. These methods add another dimension to the data by isolating ions of interest first and then subjecting them to various fragmentation methods, such as collision-induced dissociation. When coupled with liquid chromatography, precursor ions are usually selected dynamically (data-dependent acquisition); however, modern instruments allow for performing data-independent acquisition, collecting even more data about the sample.[41] *ESI-TOF-MS* and *ESI-Q-TOF-MS* provide much more data about the studied system (e.g., catalytic or biological system); moreover, the latter can provide information about the structure of the studied compounds while having higher resolution than *ESI-QQQ-MS* instruments. These instruments are present as few per department and therefore are suitable for collecting large datasets by potentially providing higher throughput. *ESI-Orbitrap-MS* instruments show high or ultrahigh resolution (depending on the exact model) and are also available. *ESI-FT-ICR-MS* (and *MALDI-FT-ICR-MS*) instruments generate vast amounts of data and provide excellent compound detection opportunities. Nevertheless, unfortunately, the number of instruments is quite limited (can be estimated as a few per region because of prohibitively high price[42,43] and complexity of operation,[44] see Supporting Information, "Number of instruments" section), which reduces the potential applicability of ML here. Moreover, even with the extremely high resolution of FT-ICR-MS, isomers are indistinguishable, and therefore, other separation methods (LC, GC, or IMS) are required. Altogether, as a rough estimate, mass spectrometers around the world may collect multiple petabytes of data yearly.[25]

Additionally, other methods can be coupled with mass spectrometry to provide additional dimensions to the data. One of the most prominent examples is ion mobility mass spectrometry,[45] where ions can be separated based on mobility values that depend on various characteristics, including ion shape and size.

It is important to mention that large amounts of data were already collected (including public spectra repositories[46–48]), and the corresponding MS spectra have been recorded and are currently being kept on computer storage. We argue that many new scientific discoveries are already within those data but have a risk of disappearance. All or even a majority of the spectra in the world cannot be analyzed manually but can be analyzed with machine learning algorithms; currently, large tech projects process tens of petabytes of data per day.[49] However, such an undertaking would require high-quality metadata for every spectrum and poses another major problem for research infrastructure, primarily because, currently, researchers are not willing to extensively document the origin and prior history of every sample. The development of more advanced and user-friendly electronic lab notebooks could potentially solve this problem.

The increasing focus on the amounts of collected data, not just the search for specific compounds, results in newer requirements for the analytical instruments. The required extensive level of characterization can only be achieved with methods allowing ionization of a larger set of compounds, meaning more diverse molecules at wide concentration ranges, providing high resolution to differentiate them and allowing for structure elucidation experiments. We expect that once these requirements are met, they will result in a significant transformation in the field of machine learning for mass spectrometry.

# Representative MS applications of ML

From sample collection to the presentation of results, the researcher usually goes through multiple steps, including sample preparation, data acquisition, data preprocessing, and finally, data analysis (**Fig. 3**). ML can significantly improve the quality of results and make the steps more efficient. In contrast to traditional ML algorithms, deep learning models show higher flexibility in output modalities and therefore may be preferred in some cases, such as protocol generation and structure determination. For some problems, such as quantification, the use of complex DL models may be excessive.

## Sample preparation

Reproducibility is an important problem in MS[50,51] and is commonly followed for analytical chemistry in general. With increased throughput,[52] this is an even more important problem. Even if no automation is in place, adaptive generation of sample preparation protocols could improve the quality of the results. Moreover, to reduce the number of errors, quality control measures should be put in place. This includes outlier detection and automated comparison of spectral data for multiple attempts.[53] Some instrument parameters could be set incorrectly and were not noticed by the researchers. This is a large field of interest for unsupervised machine learning algorithms (**Fig. 3-1**). Such problems usually benefit the most from applying ML.

## Data acquisition

Samples for mass spectrometry studies are usually separated first to reduce ion suppression and ion competition effects, which may result in inaccurate quantification of the target compound.[17] Common strategies include gas chromatography (GC) and liquid chromatography (LC) (**Fig. 3-2**). In these methods, many parameters can be changed; for instance, for LC, mobile phase composition, column used, and temperature have the most significant effect on the results. When an operator performs these procedures, these parameters are optimized manually based on the prior knowledge of the expert. However, retention times can be predicted with machine learning.[54,55] Having a model that maps these parameters and compound information to retention time, one could optimize them to achieve the desired separation.

Mass spectrometry may require tuning and optimization; depending on the instrument, there could be many parameters for the ionization source, ion transfer (which guides and transports ions to the analyzer), and analyzer. The main task here is usually to maximize the S/N ratio and obtain the signal with higher abundance. Again, here, it is also an iterative process, where a researcher continuously optimizes the instrument's parameters.

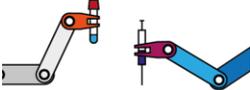

**Figure 3. Scope of potential applications of machine learning methods for mass spectrometry and related analytical chemistry fields.** Each column represents individual steps from sample preparation to analysis. At each step, key application areas are identified. By "cloud laboratories", we mean laboratories with a high degree of automation where entire protocols can be executed remotely without human intervention.

Both problems require some strategy to solve. In particular, a model could perform a specific action that would improve the final metric by reducing it in the short term. Such problems, where a model may have to perform locally suboptimal actions, are classic tasks for reinforcement learning. For example, a model could perform multiple steps of optimizing instrument parameters, even though the signal intensity may decrease at some steps. These problems usually have a non-differentiable loss function, which requires the use of specialized approaches. Unfortunately, to date, there are no examples of reinforcement learning applications to improve data collection in mass spectrometry.

## Data preprocessing
Raw data may contain many noninformative signals and a significant amount of noise (**Fig. 3-3**). Although machine learning solutions handle these situations very well,[56] some data preprocessing may be required. For chromatographic data, machine learning was applied for peak detection[57,58] and peak alignment.[59] For mass spectrometry data, deisotoping[25] and spectra deconvolution[60] may be performed. It would be interesting to see peak picking and denoising tasks being solved with machine learning. For data-independent acquisition processing of the data becomes a significant problem and can be solved with deep learning.[61]

## Data analysis
In many cases, data analysis can be performed on three levels: individual compounds, specific samples, and sets of samples (**Fig. 3-4**). Each of these levels requires the use of different algorithms and sometimes different types of preprocessing techniques.

At the individual compound level, researchers are usually interested in formula assignment and structure-solving tasks.[62] **Formula assignment** is traditionally solved with standard brute-force algorithms (sometimes using MS/MS data)[63] or via database search. However, the usage of algorithms such as neural networks[25] and gradient-boosted decision tree ensembles is possible[64] by providing significant improvements in terms of speed and quality over standard baselines.

To **solve chemical structures,** researchers usually rely on spectra obtained with destructive ionization techniques (such as electron ionization) and tandem mass spectrometry data (for instance, collision-induced dissociation, electron-transfer dissociation and electron-capture dissociation). The primary way to make conclusions from the data is through a database search.[65] When performing the search, the main question is how to measure the similarity between spectra. The naïve approach of using cosine distance showed suboptimal performance and was attempted to be replaced with machine learning ranking solutions, which resulted in increasing performance.[66–68] Additional information (such as retention times or mobility values) can improve analysis performance.[69,70] A more significant problem is that such an approach is impossible for previously unknown compounds. This is a problem where the advantages of integrating ML are the most prominent. Possible integration directions include the prediction of molecular fingerprints[71] or chemical structures directly.[72,73] The latter problem was also reformulated in terms of reinforcement learning.[74] Additionally, the problem can be mediated by solving the direct spectrum task either with quantum chemistry methods or ML.[75–77] Finally, in proteomics, specific structural features can be studied.[78,79]

One of the important challenges here is the development of datasets for a fair comparison between database searches and *de novo* molecular generation algorithms.[72,80] Moreover, current chemistry-related databases mainly comprise metabolites and other small organic molecules, disregarding the need for such datasets for transition metal complexes in catalysis. Finally, the work on using complementary spectroscopic techniques with machine learning algorithms is quite limited but would provide another dimension to the input data and increase the prediction quality.

Sometimes it is not enough to analyze a specific compound, as differences between samples may be present in the form of very complex relationships between specific distributions, making it a perfect task to show the full power of machine learning. Solutions at this level analyze **a particular sample** as a whole. For instance, ML was used to distinguish healthy and cancer tissues,[81] different types of neurons,[82] check tomato quality,[83] diagnose COVID-19,[84] and classify different types of wine[85] and oils.[86] Both deep learning and traditional ML approaches can be used here. The main challenge of current research is how such ML predictive systems could be built with a small amount of training data. This raises the need for methods for unsupervised/self-supervised learning of spectra representations, which could then be used as an input for simpler ML models, which are less likely to overfit. One of the approaches to fostering such developments is the creation of standardized benchmarks for ML tasks in mass spectrometry. Further work on public mass spectra repositories will also be important.

When **large arrays of samples** are collected, distributions of signals can provide a large amount of important information. For instance, there could be multiple groups of very similar spectra, or some compounds may tend to appear together, while others may not. Such relationships are very hard to extract with the naked eye, but such tasks are relatively simple for ML algorithms. The main task here is spectra clustering, which can be performed on top of simple spectra vectorization techniques[87] or using deep neural network embeddings.[88] In addition to large cohort studies, imaging methods (such as MALDI imaging and DESI — desorption electrospray ionization — mass spectrometry imaging) provide large datasets suitable for the first group of tasks.[89–91] Additionally, large datasets can be used to extract relationships between different analytical methods, such as NMR-MS correlation spectroscopy (SHY — statistical heterospectroscopy).[92]

# Protocol for solving general problems in mass spectrometry using ML

Solving mass spectrometry problems using machine learning involves a complex process with various challenges at each stage. In most cases, a researcher may need to consider the following aspects of the overall problem: data sources, spectra representation, metric selection, and algorithm choice. It is important to note that some aspects may be more important than others, depending on the particular problem.

**Data source**

The first question to answer is what data sources will be used (**Fig. 4a**). There are two main types of data sources: experimental and synthetic. Experimental data may be too expensive to collect but represent raw data distribution in the best way. Typically, projects involving experimental data analysis require collaboration between experimental chemists and computational scientists. One of the main challenges in this area is aligning the interests of both parties. Providing additional training for experimental chemists on the type of data and the format expected for ML work would be helpful (for example, preserving raw data or using machine-readable experiment descriptions).

One of the main advantages of mass spectrometry is an excellent agreement between theoretical and experimental masses and isotopic distribution (importantly, it depends on a specific method and experimental design). It opens a wide range of opportunities to generate synthetic data. Some research is focused on how these synthetic spectra can be augmented to match the distribution of real data.[25,93] Augmentation of real data to improve its applicability is also described.[94] To date, generative adversarial networks have no application to create synthetic data (although they were used to sample corresponding structures[95,96]), but the approach looks promising.[96] Autoregressive language models such as GPT[97] could also be applied in the mass spectrometry domain by representing a spectrum as a sequence, training the model to produce these sequences, and then sampling newer spectra from the model. However, these models are known to hallucinate in natural language generation applications[98], and their impact has yet to be explored. The use of synthetic data can potentially impact other parts of the pipeline, for example, by eliminating certain preprocessing steps during training.

Another important question to answer is whether labeled data are available (**Fig. 4a**). For many sample-oriented problems, it is usually not a problem, as detailed information about the sample is already present. However, the labels may be costly to obtain on an individual compound or array of sample levels. In this case, it is generally recommended to stick to synthetic data generation or unsupervised machine learning algorithms such as commonly used clustering algorithms (for example, k-means clustering).

Currently, challenges related to data include operation in low-data regimes, use of synthetic data, experimental method selection and optimization. Due to the high cost of experimental data, it is natural to want to collect as little data as possible, making methods for working with such amounts of data increasingly important. There is still no systematic description of how synthetic data could be generated and what could be used to make it more realistic. Finally, the use of ML in the experimental side of work also presents challenges. Any potential work on intelligently selecting and optimizing experimental methods has many chances to significantly transform the field.

**Spectra representation**

Before running an actual machine learning algorithm, mass spectra should be properly represented (**Fig. 4b**). First, the spectrum can be used as an input to another object extraction algorithm.[13] Among such objects could be chemical structures or molecular formulas. Then, further representations are derived from the corresponding set of objects. For instance, chemical structures could be used to compute molecular fingerprints, statistics over which would be computed next, and finally, used as an input to the machine learning algorithm.

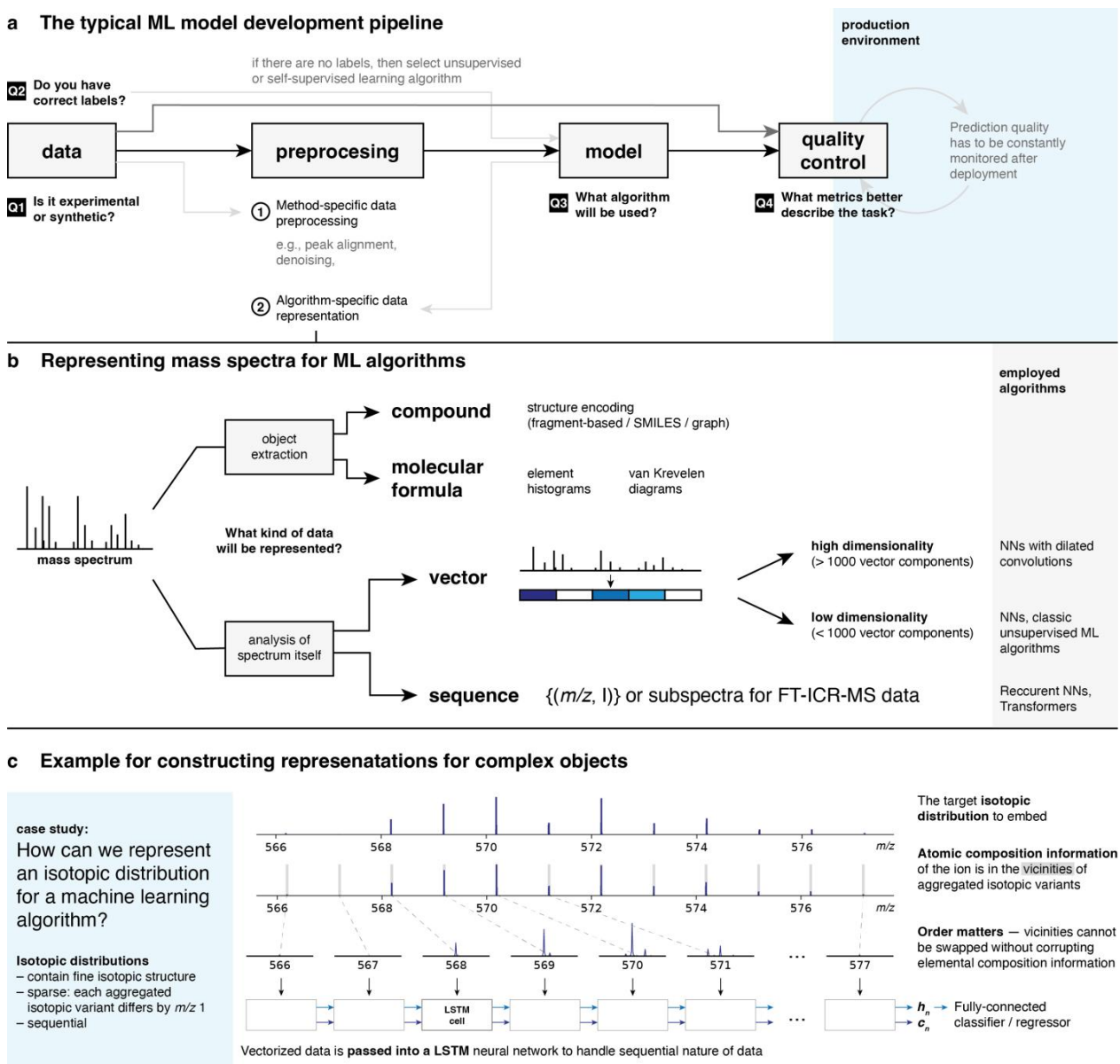

**Figure 4. Practical considerations for using ML algorithms for mass spectrometry data processing.** a — the typical ML model development pipeline (includes data collection, its preparation, and training of the model, followed by quality control); b — the ways to represent a mass spectrum to an ML algorithm; c — example for representing complex objects

The second, more widely used solution is to represent the spectrum itself. In general, it can be described in two ways: as a fixed-length vector or as a set of mass-intensity pairs. Fixed-length vectors are a common data type for many machine learning algorithms; however, when a simple binning operation is used, some important information about masses can be lost, as binning usually results in losing mass determination precision, which is important for distinguishing different ion compositions. One could prefer to use high-dimensional vector representations to preserve this information, but this limits what algorithms can be used under a fixed number of data points, as higher input data dimensionality usually implies more parameters for the model (for linear regression, the number of parameters will be equal to the dimensionality of the vector plus one), which in turn can result in overfitting. A sequence of mass-intensity pairs or subspectra (in the case of fine isotopic structure, see **Fig. 4c**) is a much more natural approach (as it can operate over input data directly, without a featurization step) but requires the use of algorithms capable of working with variable-length sequences (such as recurrent neural networks, long short-term memory networks,[99] or Transformers[100]).

In data representation, one problem particularly stands out: the confirmation of chemical structures. It is usually done based on fragmentation data. For electron ionization spectra, the data are highly reproducible not only within a single instrument model but also across different manufacturers. The reproducibility is so remarkable that a simple vector-based search works well. However, a significant amount of data is collected using tandem mass spectrometry (MS/MS) experiments, where spectra can vary significantly between instruments and depend greatly on recording parameters. Developing methods to map these different spectra of the same ion to the same point in chemical space would dramatically improve our ability to extract structural information from MS/MS experiments.

Current challenges related to spectra representation include the following issues. Is there a universal representation capable of reducing disparities between different experimental methods and instrument parameters? If there is, then it is crucial to have such datasets where invariance to data collection methods can be confirmed. Should we focus on sequence or fixed-length representations? This question is primarily related to algorithm development, as if we had effective approaches for working with long sequences, it would be more likely for us to choose them.

**Algorithm development**
The choice of algorithm is influenced by the approach used for spectral representation. However, an overview of applications reveals that only a limited number of them have been applied to these problems. Specifically, the literature on the following aspects is limited. First, the use of Transformer-based models on mass spectrometry data directly appears promising, given their success with other types of data.[101–103] Additionally, employing reinforcement learning to address complex problems with nondifferentiable metrics, such as experimental method optimization, could be a viable strategy. Last, the simultaneous combination of multiple modalities seems promising, for example, by considering chemical structure and mass spectra at the same time or by utilizing different analytical methods, such as combining mass spectrometry and spectroscopy data.

Another potential approach involves further abstraction from individual research problems and utilizing large language models (LLMs) to carry out the scientific process itself. Initial indications in this direction emerged after the release of GPT-4 by OpenAI.[103,104] It was demonstrated that these models can work with multiple types of data, suggesting that future work for scientific applications could involve the development of multimodal LLMs for various kinds of chemical information, including mass spectra.

Current challenges related to algorithm development include multimodal learning approaches, efficient feature extraction, the exploration of various algorithm choices, and the use of LLMs. The ability to combine multiple types of data is crucial for more advanced applications of ML for mass spectrometry, as it is usually not the only method applied to study a certain system. The feature extraction issue is closely related to the representation problem, but in this case, it is supposed to be learnable. As the number of MS benchmarks is limited, there is no clear understanding of which algorithms perform better for each problem. Finally, the use of LLMs is in its early days, and even potential applications have to be explored.

**Metric selection**

Unfortunately, the metric selection problem is often overlooked and does not receive enough attention. As the field receives more attention, nonexperts increasingly utilize readily available technologies. Although this mostly has a positive impact and fosters scientific development, it is also where mistakes are likely to occur.[105] One typical example is when researchers choose accuracy as the main target for imbalanced classification problems (for example, a diagnostic kit for a rare disease would have nearly perfect accuracy by giving negative outputs all the time). Fortunately, this issue can be easily addressed by developing clear guidelines that cover a wide range of potential problems.

A current challenge is the development of recommendations for proper evaluation metric choices for different applications. Scientific publications could adopt requirements for metric reporting. For instance, they could require reporting support for classification metrics. If authors report AUROC, then the corresponding curve could be required too. Researchers often attempt to tackle even more complex challenges, such as ranking, where the metrics involved can be quite complex. Moreover, these guidelines should include recommendations allowing to avoid pitfalls such as data leaks.[106]

# Conclusions and outlook

Modern mass spectrometry is in a time of great transformation, empowered with machine learning techniques, and it can be confidently said that we are entering a new era in mass spectrometry. As a result, the research community can soon anticipate a quantum leap in mass spectrometry applications. This may affect many related fields (from biochemistry to materials science), enabling their development in the future.

Furthermore, these changes can already be observed. Currently, machine learning helps us to use mass spectrometry as a diagnostic tool for various diseases and to determine the structures of unknown compounds (including *de novo* protein sequencing). Moreover, machine learning is improving mass spectrometry itself by providing tools for data acquisition and preprocessing.

Critical analysis shows that newer machine learning approaches require increasingly more data, increasing the need for automation in mass spectrometry at the steps of sample preparation, data acquisition, and data preprocessing. Moreover, not only the number of recorded samples but also the dimensionality of the obtained data should be increased, which means the use of higher resolution instruments and making these instruments more accessible for researchers. For instance, less expensive versions of FT-ICR-MS, Orbitrap or similar resolution/data-scale instruments would have a transformative effect on the field. Algorithms capable of using data from readily available lower resolution instruments would benefit the research community greatly as well.

Representation of mass spectra is still a significant challenge. For now, it appears that there is no clear understanding of how spectra at various resolutions should be represented for machine learning algorithms. Moreover, the problem of different data obtained on different instruments (particularly important for tandem mass spectra) is not properly addressed. Given recent advancements in natural language processing, these problems are likely to be solved by deep neural networks trained to represent such data in an unsupervised way fully employing already collected data.

Another significant potential direction for future research is the utilization of large language models (LLMs). Researchers may focus on using them for experimental design, method optimization, and choosing the best analysis approach based on literature data. The development of multimodal models capable of operating on mass spectrometry data could transform our approach to conducting research in the chemical sciences.

Finally, for ML in MS research to truly advance, it should focus on being more reusable for other applications. For example, instead of applying a clustering algorithm to a specific group of mass spectra, a more generally applicable tool should be developed (for example, models that are capable of representing a diverse set of spectra and are invariant to the conditions of spectra recording and instrument types). This highlights the importance of developing frameworks for analyzing MS data, having proper guidelines on open science, publishing the code, and ensuring that the code's structure allows it to be easily applied to new problems. Moreover, to further foster developments in this field, a set of benchmarks should be created to enable proper comparison of various approaches across numerous tasks.

## Competing interests

## Author contributions

## Acknowledgements
The authors thank Dr. Dmitry B. Eremin, Dr. Julia V. Burykina, and Valentina V. Ilyushenkova for helpful feedback.